\documentclass{article}

\usepackage[nonatbib,final]{neurips_2023_ml4ps}

\usepackage[utf8]{inputenc} % allow utf-8 input
\usepackage[T1]{fontenc}    % use 8-bit T1 fonts
\usepackage{hyperref}       % hyperlinks
\usepackage{url}            % simple URL typesetting
\usepackage{booktabs}       % professional-quality tables
\usepackage{amsfonts}       % blackboard math symbols
\usepackage{nicefrac}       % compact symbols for 1/2, etc.
\usepackage{microtype}      % microtypography
\usepackage{xcolor}         % colors
\usepackage{graphicx}
\usepackage{caption}
\usepackage{subcaption}

\usepackage[T1]{fontenc}    % use 8-bit T1 fonts
\usepackage{url}            % simple URL typesetting
\usepackage{booktabs}       % professional-quality tables
\usepackage{amsfonts}       % blackboard math symbols
\usepackage{nicefrac}       % compact symbols for 1/2, etc.
\usepackage{microtype}      % microtypography
\usepackage{subcaption}
\usepackage{notoccite}
\usepackage[cmex10]{amsmath}
\usepackage{amsthm}
\usepackage{supertabular}
\usepackage{times}
\usepackage{amsthm}
\usepackage{floatrow}
\usepackage{wrapfig}
\usepackage{tikz}
\usepackage{float}
\usepackage{mathtools}
\usepackage{textcomp}
\usepackage{multirow}
\usepackage{algpseudocode}
\usepackage[ruled,vlined]{algorithm2e}
\usepackage{adjustbox}
\usepackage[justification=centering, font=small]{caption} 
\usepackage{epstopdf}
\usepackage{tabularx}
\usepackage{nccmath}
\usepackage{mathtools}
\usepackage{easyReview}
\usepackage{tikz}
\usepackage[justification=centering, font=small, skip = 0pt]{caption} 
\usepackage{epstopdf}
\usepackage{tabularx}
\usepackage{easyReview}
\usepackage{comment}

\usepackage{pdfpages}

\theoremstyle{plain}

\numberwithin{corr}{section} 

\numberwithin{defn}{section}

\numberwithin{prop}{section} 
\newcommand{\vect}[1]{\mathbf{#1}}

\graphicspath{{./Figures/}}

\title{Forward Gradients for Data-Driven CFD Wall Modeling}

\author{%
  Jan H\"uckelheim \\
  Argonne National Laboratory \\
  Lemont, IL, USA \\
  \texttt{jhueckelheim@anl.gov}
  \And
  Tadbhagya Kumar \\
  Argonne National Laboratory \\
  Lemont, IL, USA \\
  \texttt{tkumar@anl.gov}
  \AND
  Krishnan Raghavan \\
  Argonne National Laboratory \\
  Lemont, IL, USA \\
  \texttt{kraghavan@anl.gov}
  \And
  Pinaki Pal \\
  Argonne National Laboratory \\
  Lemont, IL, USA \\
  \texttt{pal@anl.gov}
}

\begin{document}

\maketitle

\begin{abstract}
    Computational Fluid Dynamics (CFD) is used in the design and optimization of gas turbines and many other industrial/ scientific applications. However, the practical use is often limited by the high computational cost, and the accurate resolution of near-wall flow is a significant contributor to this cost. Machine learning (ML) and other data-driven methods can complement existing wall models. Nevertheless, training these models is bottlenecked by the large computational effort and memory footprint demanded by back-propagation. Recent work has presented alternatives for computing gradients of neural networks where a separate forward and backward sweep is not needed and storage of intermediate results between sweeps is not required because an unbiased estimator for the gradient is computed in a single forward sweep. In this paper, we discuss the application of this approach for training a subgrid wall model that could potentially be used as a surrogate in wall-bounded flow CFD simulations to reduce the computational overhead while preserving predictive accuracy. 
    % We implemented this approach in a framework built on top of PyTorch and
\end{abstract}

\section{Introduction}
%Describe the science problem. 
%What benefits will one have by leveraging the AI system for the science?
Reverse mode automatic differentiation and back-propagation can efficiently compute gradients, but often require large amounts of memory. This is due to the need for storing intermediate state information. Moreover, these computations are not supported on all hardware, particularly on some AI accelerators. In this work, we explore the use of forward gradients as an alternative to back-propagation in the context of data-driven machine learning models to augment conventional computational fluid dynamics (CFD) methods for simulating film cooling of turbo-machinery components such as gas turbines.

To perform parametric design studies of such components, it is often desired to have high-fidelity simulations that capture a wide range of spatio-temporal scales. Wall-resolved large-eddy simulations (WRLES) \cite{liu2005large, guo2006large, peet2008near, oliver2017implicit, nunno2022wall} capture the unsteady features and provide a feasible way for simulating complex turbulent flows at a reasonable cost. However, the computational cost incurred due to very high resolution needed to resolve the viscous scales near the wall is still high. To overcome this, wall models are used to avoid the need to resolve the near-wall region, providing a feasible way for LES of wall-bounded flows at high Reynolds numbers.

Recent advancements in machine learning and high performance computing have motivated new efforts to develop data-driven methods that complement existing wall models. Some popular methods and architectures in the literature that facilitate such development are Random Forest regression (\cite{wang2017physics, tkumar}), artificial neural networks {\cite{milano2002neural, zhou2021wall, yang2019predictive, huang2019wall}}, and convolutional neural networks{\cite{kim2020prediction, guastoni2020prediction, https://doi.org/10.48550/arxiv.2106.09271}}. In most of these approaches, high fidelity WRLES are performed to gather data corresponding to flow features~(velocity of the fluid, pressure gradients, etc.). Then, a data-driven regression model is trained to predict the wall-shear stress from these flow features, which is used as a wall model for LES of turbulent channel flows with coarser near-wall resolution.

\section{Application}
In this application, we seek to develop a data-driven wall model to predict wall shear stress in large-eddy simulation (LES) of a gas turbine film cooling configuration. Film cooling is a popular technique used to reduce turbine temperature and protect them from thermal failure. As the hot gases exit from the combustor and enter the turbine stage, the temperatures can be higher than the melting point of turbine stage materials. Cooling strategies thus become necessary for thermal management of the turbine blades. Since conventional CFD models suffer from the aforementioned limitations, neural network-based wall models are being investigated as a surrogate in wall-bounded flow simulations to reduce the computational overhead while preserving predictive accuracy.

\subsection{Data Generation}
\begin{figure}
    \centering
    \includegraphics[width=0.9\linewidth]{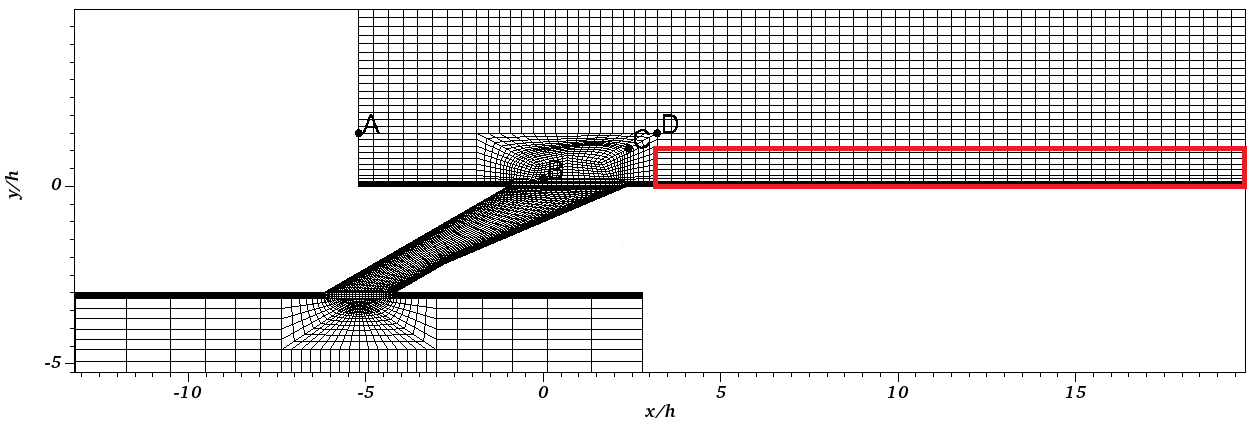}
    \caption{Wall resolved LES mesh of the film cooling setup. The highlighted region (red) depicts the domain of data collection for training the ML wall model}
    \label{mesh}
\end{figure}

The data used in this work to train the data-driven model is derived from WRLES of a 7-7-7 cooling hole configuration detailed in \cite{nunno2022wall} for two blowing ratios (BR): 1 and 1.5. The film cooling configuration consists of three parts: a plenum feeding cool air to a flat surface via a single row of cooling holes (represented with a single hole with periodic boundary conditions in the spanwise direction). The flow was simulated using the Nek5000 \cite{nek5000} platform, a higher-order spectral element CFD code, using its Low Mach flow solver. Figure \ref{mesh} depicts the WRLES mesh used to perform the CFD simulation, and the red box highlights the flow domain which is used to train the data-driven model. The flow data in the highlighted region was collected at 91 planes for $y \in [0.05, 0.5]$ such that $z \in [-3,3]$ and $x \in [3.2, 19.6]$.

\subsection{Data-driven Modeling}
\begin{figure}
    \centering
    \includegraphics[width=0.6\linewidth]{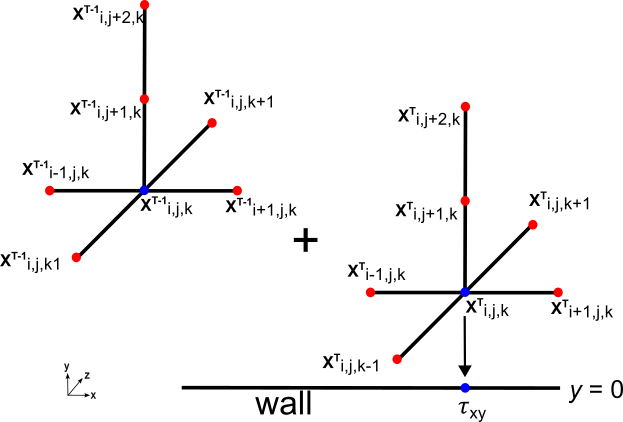}
    \caption{Combining features in a stencil form at two timew, $t = T$ and $t = T-1$ to make streamwise shear stress predictions ($\tau_{xy}$) at a point. $\textbf{X}$ denotes the feature set at a point consisting of velocity and velocity gradients, the superscipt denotes the time, and the subscripts denote the coordinates.}
    \label{stencil}
\end{figure}

Fluid velocity and velocity gradients extracted from the WRLES serve as the Ml model input features, whereas the streamwise shear stress is used as the output. To make shear stress predictions at a grid point located at $(i,j,k)$, pointwise flow features from neighboring x, y, z locations are combined in a stencil-like form. If the pointwise feature set consisting of fluid velocity components and gradients is denoted by $\textbf{X}_{i,j,k}$, the corresponding 3D stencil feature set is given by:
\begin{equation}
    \textbf{X}_{st} = \textbf{X}_{i,j,k} + \textbf{X}_{i-1,j,k} + \textbf{X}_{i+1,j,k} + \textbf{X}_{i,j,k+1} + \textbf{X}_{i,j,k-1} + \textbf{X}_{i,j+1,k} + \textbf{X}_{i,j+2,k} 
\end{equation}
The feature stencil at time $t = T$ is combined with stencil features with time delay ($t = T - 1$) to form the input features to the network, as shown in Figure \ref{stencil}. The output label is the streamwise shear stress ($\tau_{xy}$) at time $t = T$. The network consists of an input layer of size equal to the number of features used and 3 hidden layers with 64 neurons each with a dropout value of 0.1 and ReLU activation function. The output layer is a dense linear layer of size one and predicts the wall shear stress in the streamwise direction ($\tau_{xy}$). The prepared input - output data is scaled in the range [-1,1] using Min-Max scaling. The dataset for BR1 is randomly shuffled and an 80\%-20\% split is performed to generate training and validation datasets, respectively. The loss function used to train the network is the mean squared error ($E_w$):
\begin{equation}
    E_w = \frac{1}{N} \sum_{i=1}^{N} (y_i - \hat{y}_i)^2
\end{equation}
The neural network model is implemented in PyTorch and is trained using the Adam optimizer with early stopping criterion. % The trained model performance is assessed using the Pearson’s correlation coefficient of the predicted wall shear stress against the WRLES data on our validation set.% Additionally, the model is tested on BR1.5 flow data to ascertain the generalization and extrapolation capabilities. TODO do we actually do this here?
 %% Ted - No it isn't done here.
\section{Forward Gradients}
 The key contribution of this paper is the implementation and use of forward gradients for a physical science use-case, not demonstrated earlier. We summarize the theory as shown in~\cite{baydin2022gradients}, followed by a description of our implementation strategy.

\subsection{Theory}
\label{sec:theory}
Given a objective function $f(\vect{\theta}):\mathbb{R}^{n} \rightarrow \mathbb{R}^{n},$ we seek to solve the optimization problem
\begin{align}
    \underset{ \vect{\theta} \in \vect{\Omega} }{min} f(\vect{\theta}).
\end{align}
Intuitively, we seek to find a $\vect{\theta}$ in the search space  
$\vect{\Omega}$ such that the objective function value is minimized. The objective function can take many forms but is usually expected to be twice differentiable. This is required because, typically, the process of finding the minima involves a search using the gradients of the objective function denoted 
$ \nabla_{\vect{\theta}} f(\vect{\theta}).$ Since exact calculation of $ \nabla_{\vect{\theta}} f(\vect{\theta}) $ is usually expensive, several simplifications are performed based on the assumption that, what is usually required for efficient learning is the directional derivative of $f$ with respect to $\vect{\theta}.$ This can be accomplished by forward gradients~\cite{baydin2022gradients}  $g(\vect{\theta})$ such that
\begin{align}
    g(\vect{\theta}) = (\nabla_{\vect{\theta}} f(\vect{\theta}) . \vect{v}) \vect{v}
    \label{eqn:optim}
\end{align}
where $\vect{v}$ is a perturbation vector chosen as a multivariate random variable sampled $\vect{v} \sim p(\vect{v})$ such that the scalar components $v_{i}$ are independent with zero mean and  unit variance for all $i'$s. The key insight here is that, $g(\vect{\theta}) $ through the perturbation vector allows us to estimate the directional derivatives without needing to explicitly compute the gradient $\nabla_{\vect{\theta}} f(\vect{\theta}).$ This allows for a runtime advantage over typical back-propagation where we can interpret the directions of update without needing to distinguish the updates for each scalar parameter in $\vect{\theta}.$
The process of evaluating forward gradients typically involves a three step process  where we
\begin{enumerate}
    \item sample a random perturbation $\vect{v} \sim  p(\vect{v})$, which has the same size with $f'$s argument,
    \item run forward mode autodiff to evaluate $(\nabla_{\vect{\theta}} f(\vect{\theta}) . \vect{v})$ and $ f(\vect{\theta})$ simultaneously, and finally 
    \item multiply  $(\nabla_{\vect{\theta}} f(\vect{\theta}) . \vect{v})$ with the perturbation vector $\vect{v}$ to obtain the forward gradients $g(\vect{\theta}).$
\end{enumerate}

As long as $g(\vect{\theta})$ is similar to the true gradients, or points in a descent direction, the optimization problem can still be solved efficiently. It is shown in~\cite{baydin2022gradients} that the forward gradients are an unbiased estimator of $\nabla_{\vect{\theta}} f(\vect{\theta})$ and can be used to solve the optimization problem~\eqref{eqn:optim}.

\subsection{Implementation}
Forward gradients are propagated through the model using the forward mode, also known as tangent mode, of automatic differentiation~\cite{griewank2008evaluating}. Since the code used by~\cite{baydin2022gradients} is not yet publicly available at the time of writing, we used our own framework, which not only supports the computation of randomized forward gradients, but also computes entire Jacobians of the network or individual layers, as well as mixed-mode differentiation to freely combine forward gradients with conventional back-propagation across layers. The implementation is already freely available on github\footnote{URL removed for anonymization}.

The framework implements a so-called \emph{tangent model} for selected PyTorch layers, which implement the forward mode of automatic differentiation for these layers. It also provides a simple way to combine multiple layers in a way that mimics standard PyTorch usage, and can therefore be used as a drop-in replacement for PyTorch in many cases. The tangent model of individual layers is implemented by using existing PyTorch layers underneath, which allows us to support GPUs, AI accelerators, and other hardware platforms efficiently. Since our framework can compute forward gradients without using PyTorch back-propagation, it can even be used to obtain approximate gradients on AI accelerators.
% that do not officially support training due to their lack of back-propagation support.

We discuss here the tangent model of a 2D convolution layer to illustrate the method. For each output pixel and output channel $N_i$ and $C_{{out}_j}$, \texttt{Conv2d} computes an output value $o$ by adding a bias to a weighted sum of cross correlations over the input values $u$ at nearby pixels across input channels as
\begin{equation}
{o}(N_i,C_{{out}_j}) = {b}(C_{{out}_j}) + \sum\limits_{k=0}^{C_{in}-1}{w}(C_{{out}_j},k)\star{u}(N_i,k).
\end{equation}
Differentiating with respect to the bias, weights and inputs yields
\begin{equation}
\dot{{o}}(N_i,C_{{out}_j}) = \dot{{b}}(C_{{out}_j}) + \sum\limits_{k=0}^{C_{in}-1}\dot{{w}}(C_{{out}_j},k)\star{u}(N_i,k) + \sum\limits_{k=0}^{C_{in}-1}{w}(C_{{out}_j},k)\star\dot{{u}}(N_i,k),
\label{eqn:tangent}
\end{equation}
where derivative terms are denoted using a dot above. To compute forward gradients, the term $\dot{w}$ needs to be replaced with a random vector as discussed in Section~\ref{sec:theory}. The term $\dot{u}$ needs to be replaced with the output of the preceding tangent layer, or dropped in case of the first layer of the network. The tangent model shown in \eqref{eqn:tangent} can itself be computed using two calls to \texttt{Conv2d}.
 
\section{Evaluation}
\begin{figure}
     \centering
     \begin{subfigure}[b]{0.49\textwidth}
         \centering
         \includegraphics[width=\textwidth]{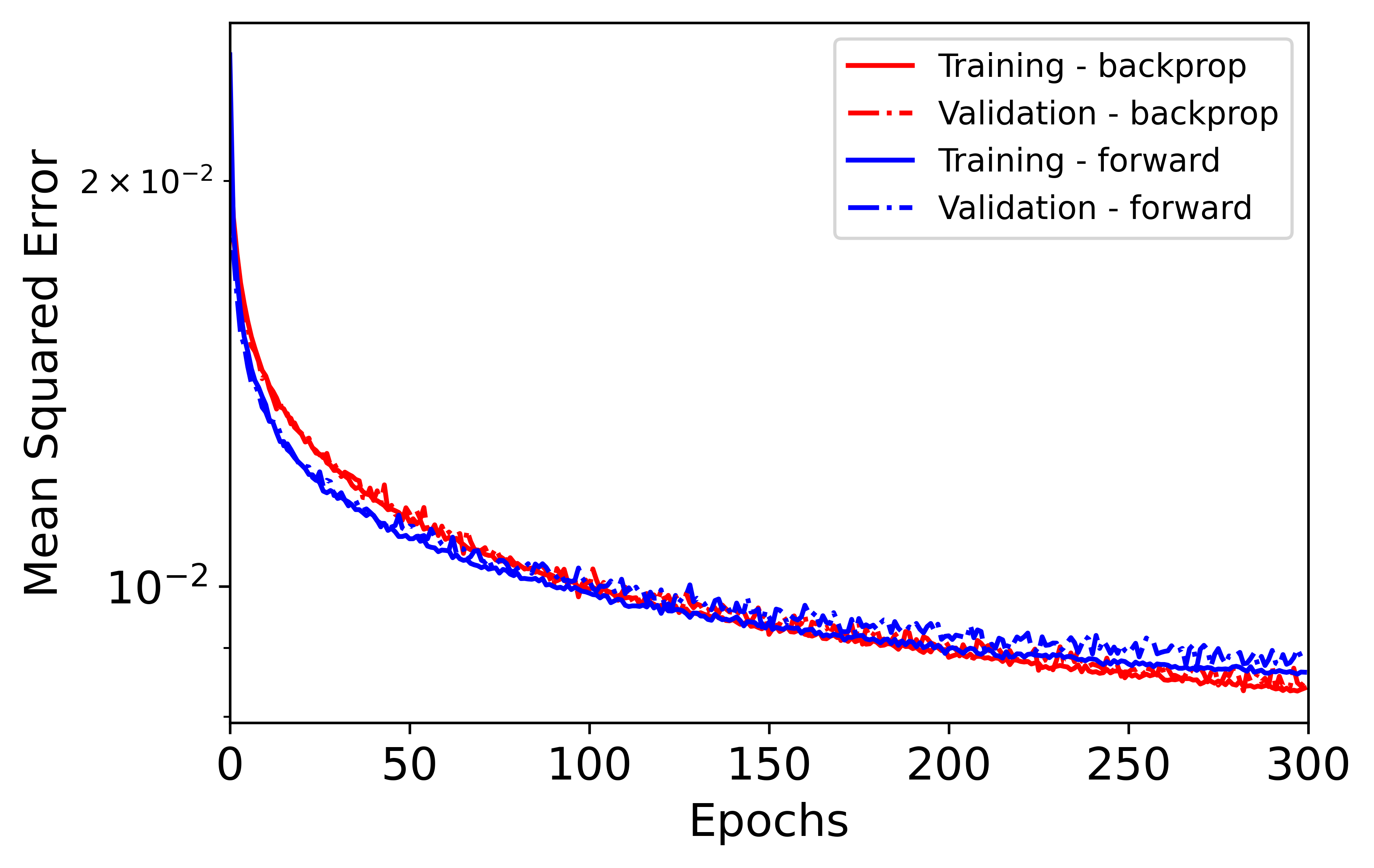}
         \caption{}
         \label{loss_plot}
     \end{subfigure}
     \hfill
     \begin{subfigure}[b]{0.49\textwidth}
         \centering
         \includegraphics[width=\textwidth]{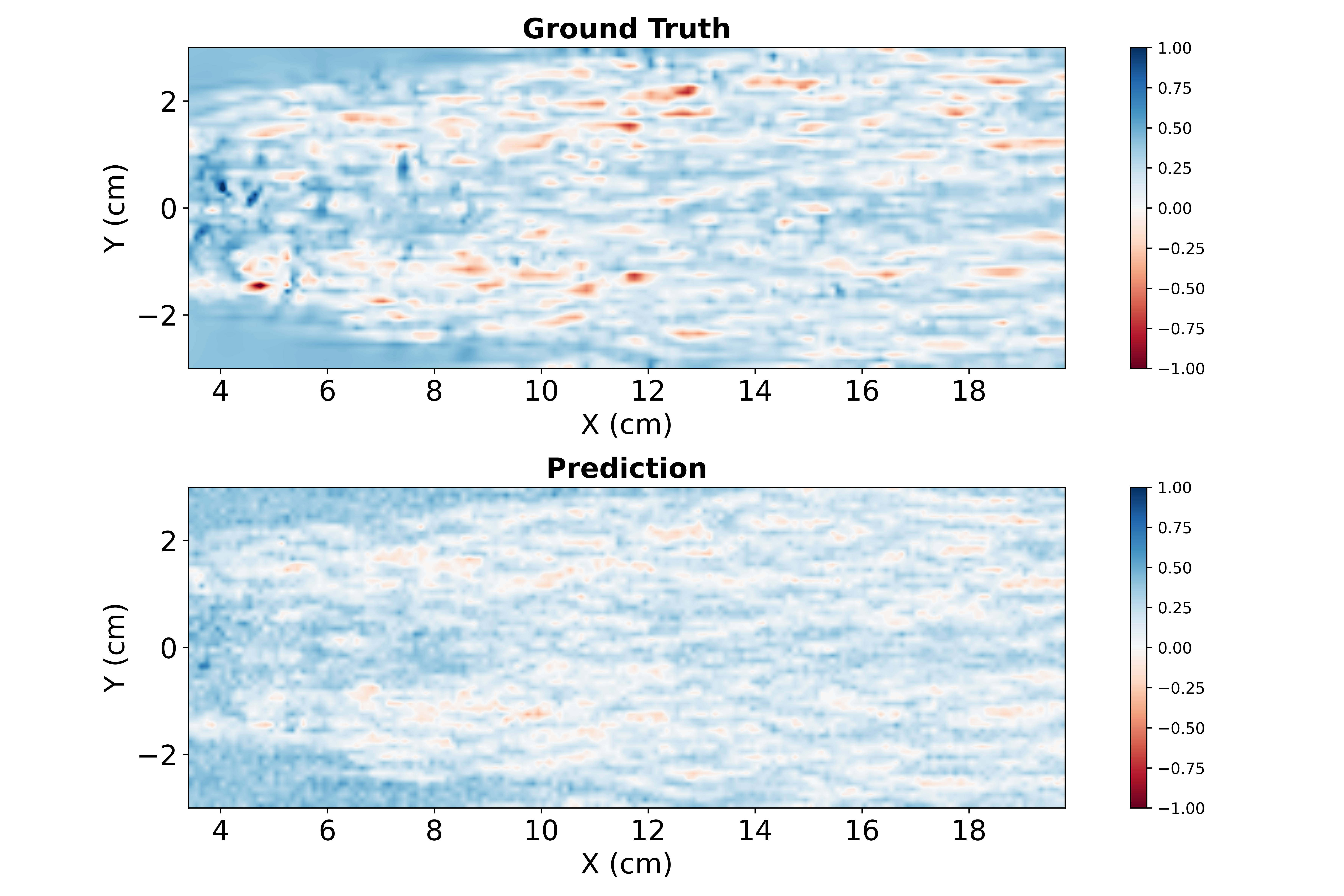}
         \caption{}
         \label{preds}
     \end{subfigure}
        \caption{Comparison of a) loss evolution between the forward gradient approach and standard back-propagation for training on the wall-resolved LES dataset, and b) out-of-sample predictions for normalized instantaneous wall shear stress from the forward gradient approach versus the ground truth.}
        \label{fig:}
\end{figure}
Figure \ref{loss_plot} shows the loss function evolution for both the forward gradient approach and back-propagation approach. The network is trained using ADAM optimizer, with an initial learning rate $\eta = 2 \times 10^{-4}$. As training progresses, the learning rate is decayed by a factor $\gamma = 0.5$ if the validation loss does not change for 20 epochs. Figure \ref{preds} compares the instantaneous shear stress predictions from the neural network trained with forward gradient approach against the ground truth at a time instant out of the training sample. It is seen that the trained network captures the overall shear stress distribution in the domain.  

\section{Conclusion and next steps}
We demonstrated that forward gradients can be used effectively for training without back-propagation in the context of computational fluid dynamics applications. In future work, we plan to investigate different optimizer settings and network architectures, which may lead to setups that are more suited to the randomized nature of forward gradients, and perform even better or more robustly. We also plan to experiment with computing the mean from multiple forward gradient computations with varying random seeds, which is likely to yield a more accurate representation of the true gradient at a higher computational cost, but still without the memory footprint of conventional back-propagation.

The generation of (pseudo-)random numbers is a performance bottleneck in our current implementation, since the dense layers have a large number of parameters, each of which requiring its own random number that will be subsequently used only for a small number of floating point operations. This may be a fundamental challenge with the forward gradient approach that was not discussed in~\cite{baydin2022gradients}, and provides further motivation for exploring other neural network architectures that have fewer parameters that get re-used more often.

\section{Acknowledgments}
This research used resources of the Argonne Leadership Computing Facility, and was supported by Laboratory Directed Research and Development (LDRD) funding from Argonne National Laboratory, provided by the Director, Office of Science, of the U.S. Department of Energy under Contract No. DE-AC02-06CH11357.
The gas turbine film cooling datasets were generated under a research project in collaboration with Raytheon Technologies Research Center (RTRC) funded by the DOE Advanced Manufacturing Office (AMO) through the High Performance Computing for Energy Innovation (HPC4EI) program. Lastly, the authors would like to acknowledge the computing core hours available through the Bebop cluster provided by the Laboratory Computing Resource Center (LCRC) at Argonne National Laboratory and National Energy Research Scientific Computing Center (NERSC) Cori supercomputer for the generation of the high-fidelity simulation datasets.

%\section*{References}
\bibliographystyle{unsrt}
\bibliography{main.bib}

\begin{thebibliography}{10}

\bibitem{liu2005large}
Kunlun Liu and Richard Pletcher.
\newblock Large eddy simulation of discrete-hole film cooling in a flat plate
  turbulent boundary layer.
\newblock In {\em 38th AIAA Thermophysics Conference}, page 4944, 2005.

\bibitem{guo2006large}
X~Guo, W~Schr{\"o}der, and M~Meinke.
\newblock Large-eddy simulations of film cooling flows.
\newblock {\em Computers \& Fluids}, 35(6):587--606, 2006.

\bibitem{peet2008near}
Yulia~V Peet and Sanjiva~K Lele.
\newblock Near field of film cooling jet issued into a flat plate boundary
  layer: Les study.
\newblock In {\em Turbo Expo: Power for Land, Sea, and Air}, volume 43147,
  pages 409--418, 2008.

\bibitem{oliver2017implicit}
Todd~A Oliver, Joshua~B Anderson, David~G Bogard, Robert~D Moser, and Gregory
  Laskowski.
\newblock Implicit les for shaped-hole film cooling flow.
\newblock In {\em Turbo Expo: Power for Land, Sea, and Air}, volume 50879, page
  V05AT12A005. American Society of Mechanical Engineers, 2017.

\bibitem{nunno2022wall}
Austin~C Nunno, Sicong Wu, Muhsin Ameen, Pinaki Pal, Prithwish Kundu, Ahmed
  Abouhussein, Yulia Peet, Michael Joly, and Peter Cocks.
\newblock Wall-resolved les study of shaped-hole film cooling flow for varying
  hole orientation.
\newblock In {\em AIAA SCITECH 2022 Forum}, page 1404, 2022.

\bibitem{wang2017physics}
Jian-Xun Wang, Jin-Long Wu, and Heng Xiao.
\newblock Physics-informed machine learning approach for reconstructing
  reynolds stress modeling discrepancies based on dns data.
\newblock {\em Physical Review Fluids}, 2(3):034603, 2017.

\bibitem{tkumar}
Tadbhagya Kumar, Pinaki Pal, Austin~C Nunno, Sicong Wu, Opeluwa Owoyele,
  Michael Joly, and Dima Tretiak.
\newblock Development of a data-driven wall model for large-eddy simulation of
  gas turbine film cooling flows.
\newblock In {\em AIAA SciTech Forum and Exposition}, page 1254, 2023.

\bibitem{milano2002neural}
Michele Milano and Petros Koumoutsakos.
\newblock Neural network modeling for near wall turbulent flow.
\newblock {\em Journal of Computational Physics}, 182(1):1--26, 2002.

\bibitem{zhou2021wall}
Zhideng Zhou, Guowei He, and Xiaolei Yang.
\newblock Wall model based on neural networks for les of turbulent flows over
  periodic hills.
\newblock {\em Physical Review Fluids}, 6(5):054610, 2021.

\bibitem{yang2019predictive}
XIA Yang, Suhaib Zafar, J-X Wang, and Heng Xiao.
\newblock Predictive large-eddy-simulation wall modeling via physics-informed
  neural networks.
\newblock {\em Physical Review Fluids}, 4(3):034602, 2019.

\bibitem{huang2019wall}
Xinyi~LD Huang, Xiang~IA Yang, and Robert~F Kunz.
\newblock Wall-modeled large-eddy simulations of spanwise rotating turbulent
  channels—comparing a physics-based approach and a data-based approach.
\newblock {\em Physics of Fluids}, 31(12):125105, 2019.

\bibitem{kim2020prediction}
Junhyuk Kim and Changhoon Lee.
\newblock Prediction of turbulent heat transfer using convolutional neural
  networks.
\newblock {\em Journal of Fluid Mechanics}, 882, 2020.

\bibitem{guastoni2020prediction}
Luca Guastoni, Miguel~P Encinar, Philipp Schlatter, Hossein Azizpour, and
  Ricardo Vinuesa.
\newblock Prediction of wall-bounded turbulence from wall quantities using
  convolutional neural networks.
\newblock In {\em Journal of Physics: Conference Series}, volume 1522, page
  012022. IOP Publishing, 2020.

\bibitem{https://doi.org/10.48550/arxiv.2106.09271}
Naoki Moriya, Kai Fukami, Yusuke Nabae, Masaki Morimoto, Taichi Nakamura, and
  Koji Fukagata.
\newblock Inserting machine-learned virtual wall velocity for large-eddy
  simulation of turbulent channel flows, 2021.

\bibitem{nek5000}
Paul Fischer, James Lottes, and Henry Tufo.
\newblock Nek5000.
\newblock [Computer Software] \url{https://doi.org/10.11578/dc.20210416.29},
  jun 2007.

\bibitem{baydin2022gradients}
At{\i}l{\i}m~G{\"u}ne{\c{s}} Baydin, Barak~A Pearlmutter, Don Syme, Frank Wood,
  and Philip Torr.
\newblock Gradients without backpropagation.
\newblock {\em arXiv preprint arXiv:2202.08587}, 2022.

\bibitem{griewank2008evaluating}
Andreas Griewank and Andrea Walther.
\newblock {\em Evaluating derivatives: principles and techniques of algorithmic
  differentiation}.
\newblock SIAM, 2008.

\end{thebibliography}

% \section{Acknowledgements}
% This research used resources of the Argonne Leadership Computing Facility, which is a DOE Office of Science User Facility supported under Contract DE-AC02-06CH11357.

% %%%%%%%%%%%%%%%%%%%%%%%%%%%%%%%%%%%%%%%%%%%%%%%%%%%%%%%%%%%%

% \appendix

% \section{Appendix}

% Optionally include extra information (complete proofs, additional experiments and plots) in the appendix.
% This section will often be part of the supplemental material.

%\includepdf[pages=-]{neurips_checklist.pdf}

\end{document}